\documentclass[11pt]{article}
\renewcommand{\theequation}{\arabic{section}.\arabic{equation}}
\def\be{\begin{equation}}
\def\ee{\end{equation}}
\def\ba{\begin{eqnarray}}
\def\ea{\end{eqnarray}}
\def\nn{\nonumber}
\def\lb{\label}
\def\bb{\bibitem}
\def\A{{\cal A}}
\def\E{{\cal E}}

\begin{document}

\begin{titlepage}

\date{}

\title{
\begin{flushright}\begin{small}    LAPTH-045/26\end{small} \end{flushright} \vspace{1cm}
A wormhole with two black holes}

\author{G\'erard Cl\'ement\thanks{Email: gclement@lapth.cnrs.fr} \\ \\
{\small LAPTh, Universit\'e Savoie Mont Blanc, CNRS, F-74940  Annecy, France}}

\maketitle

\begin{abstract}
We report on a new three-parameter stationary solution of the four-dimensional vacuum Einstein equations, which describes a Lorentzian wormhole harboring two antipodal co-rotating black holes. These are hung from one of the two spacelike infinities by two semi-infinite cosmic strings with positive or negative tension. We evaluate the Komar masses and angular momenta of the two black holes, and check that these obey the usual Smarr relation, as well as the generalized first law for a system of collinear black holes and cosmic strings. While free from Misner strings, this spacetime allows closed timelike curves within a bounded volume. The metric also admits a naked ring singularity for almost all values of the parameters.
\end{abstract}
\end{titlepage}
\setcounter{page}{2}

\section{Introduction}

Lorentzian wormholes may occur as solutions to the Einstein field equations with suitable ``exotic'' sources violating the weak energy condition. The most simple example is the well-known spherically symmetric Ellis-Bronnikov wormhole solution \cite{ellis,bronnikov1973} to the theory of a self-gravitating phantom scalar field. Some wormholes, however, arise as solutions to the Einstein or Einstein-Maxwell equations without manifest exotic matter, but with line singularities \cite{gibbons2017}. The maximal analytic extension of the over-rotating Kerr metric ($a^2>M^2$) is a wormhole with a naked ring singularity \cite{hawking2011}. Other examples are special Weyl solutions generated by a cosmic string loop \cite{zipoy,bronnikov1997}, or by straight Misner-Dirac strings \cite{clement2015}.

The properties of asymptotically flat stationary black hole solutions to the Einstein or Einstein-Maxwell field equations are well known. Non-asymptotically flat black hole solutions, such as asymptotically Bertotti-Robinson black holes in Einstein-Maxwell theory \cite{ernst1976,alekseev1996} have recently attracted some interest \cite{podolsky2025, astorino2025,alekseev2025}. Other non-asymptotically flat solutions are linear dilaton black holes in Einstein-Maxwell-dilaton theory \cite{chan,newdil}.

In this paper, we report on a new stationary solution to the vacuum Einstein equations which combines two features: this corresponds to a wormhole with two locally asymptotically flat regions, and two co-rotating black holes are embedded in this wormhole environment. These two twin black holes are not connected to each other by a strut or a Misner string. On the other hand, they are connected to one of the wormhole ends by semi-infinite cosmic strings with positive or negative tension. A naked ring singularity is also present.

The wormhole solution is obtained by analytical continuation of the gravimagnetic dipole vacuum solution, first constructed in \cite{manko2009}, and discussed more fully in \cite{dinut}. This asymptotically Kerr stationary axisymmetric metric describes a system of two black holes connected by a Misner string. It has been argued that such a spacetime metric might explain the flattened out profiles of the rotation curves of spiral galaxies \cite{govaerts2023}.

In the following, we first recall briefly the main results of \cite{dinut}. We then carry out an analytical continuation, yielding the new wormhole solution. We investigate the various geometric properties of this solution. Finally, we evaluate the Komar masses and angular momenta of the two black holes, and check that these obey the usual Smarr relation, as well as a generalized first law of black hole mechanics.

\setcounter{equation}{0}
\section{The gravimagnetic dipole}

 This stationary axisymmetric metric, depending on three parameters $m$ (the half-total mass), $\alpha_+$ and $\alpha_-$ (with $\alpha_+ >\alpha_- > 0$), can be cast in the Weyl-Papapetrou form:
 \be\lb{weyl}
ds^2 = - f(dt-\omega d\varphi)^2 + f^{-1}\left[e^{2k}(d\rho^2+dz^2) + \rho^2d\varphi^2\right],
 \ee
where $\varphi$ is an angle. The metric functions $f(\rho,z)$, $\omega(\rho,z)$, $k(\rho,z)$ are derived from the Ernst potential
\be\lb{ernstpot}
{\cal E} = \frac{A-B}{A+B},
 \ee
with
\be
A = A_R + iA_I, \quad B = B_R + iB_I,
\ee
and\footnote{The functions $A$ and $B$ have been rescaled by a factor $1/m^4$, the parameter $d$ has been rescaled by a factor $1/m^2$,
and the new parameter $a$ is related to the parameters of \cite{dinut} by $ma=k\nu$.}
\ba\lb{AB1}
A_R &=& - \alpha_+\alpha_-\left[2d_+d_-(R_+R_- + r_+r_-) + (d^2+1)R_Sr_S\right] \nn\\
&& + \frac{1}2\left[d_-^2\alpha_+^2 + d_+^2\alpha_-^2\right]R_Dr_D, \nn\\
A_I &=& 2ad\left[\alpha_- R_Dr_S - \alpha_+R_Sr_D\right], \nn\\
B_R &=& - 4\alpha_+\alpha_- md\left[d_-R_S + d_+r_S\right], \nn\\
B_I &=& - 4amd\left[d_-\alpha_- R_D + d_+\alpha_+ r_D\right],
\ea
where we have put
\be\lb{da1}
d = \frac{\alpha_+^2 - \alpha_-^2}{4m^2}, \quad d_\pm = d\pm1 , \quad a^2 = m^2d_+^2-\alpha_+^2 = m^2d_-^2 - \alpha_-^2.
\ee
These functions depend on the variables $\rho \ge 0$ and $z$ through the distances
 \be\lb{Rr}
R_\pm = \sqrt{\rho^2+(z\pm\alpha_+)^2}, \quad r_\pm = \sqrt{\rho^2+(z\pm\alpha_-)^2},
 \ee
of a generic ``point'' $(\rho,z)$ to the four focal points $(\rho=0,z=\pm\alpha_+)$ and $(\rho=0,z=\pm\alpha_-)$ , and the symbols $R_S$, $R_D$, $r_S$, $r_D$ stand for the combinations
\be
R_S = R_+ + R_-,\quad R_D = R_+ - R_-, \quad r_S = r_+ + r_-,\quad r_D = r_+ - r_-.
\ee
The Weyl-Papapetrou metric functions $f$, $\omega$ and $e^{2k}$ are given in terms of $A$, $B$ and an auxiliary function $G$ \footnote{The function $G$ is related to that of \cite{dinut} by $G = [G_{\rm old} - m(A_{\rm old}+B_{\rm old})]/m^4$.} by
\be\label{metfunct1}
f = \frac{A\bar{A} - B\bar{B}}{(A+B)(\bar{A}+\bar{B})}, \quad
e^{2k} = \frac{A\bar{A} - B\bar{B}}{64d^4\alpha_+^2\alpha_-^2R_+R_-r_+r_-}, \quad
\omega = - \frac{4{\rm Im}[G(\bar{A}+\bar{B})]}{A\bar{A} - B\bar{B}},
 \ee
where
\be
G = G_R + iG_I,
\ee
with
\ba \lb{G1}
G_R &=& - 2\alpha_+\alpha_- md\left[d_-(\alpha_+R_D - zR_S) + d_+(\alpha_- r_D - zr_S)\right] \nn\\
&& + m^2d\left[d_+^2\alpha_- R_Dr_S - d_-^2\alpha_+R_Sr_D\right], \nn\\
G_I &=& 2amd\left[- \alpha_+\alpha_-(d_-R_S + d_+r_S) \right. \nn\\
&& \left. + 2mdR_Dr_D + (d_-\alpha_- R_D + d_+\alpha_+r_D)z\right].
\ea

The four focal points divide the $z$-axis ($\rho=0$) in five rods. The metric (\ref{weyl}) is regular
($\omega=0$, $e^{2k}=1$) on the semi-infinite outer rods $z>\alpha_+$ and $z<-\alpha_+$.
The two rods $\alpha_-<z<\alpha_+$ and $-\alpha_+<z<-\alpha_-$ correspond to two co-rotating black hole horizons
with equal masses and opposite NUT charges. And the central rod $-\alpha_-<z<\alpha_-$ corresponds to a spinning cosmic string, or Misner string, carrying the gravimagnetic flux between the two horizons. The tension of this cosmic string can vanish
($e^{2k}=1$) in a two-dimensional subset of the three-parameter domain. The evaluation of the asymptotic behavior for large $\rho$ and $z$ shows that the metric is asymptotically flat, with total mass $M=2m$ and total angular momentum $J = Ma$ \cite{manko2009}.

\setcounter{equation}{0}
\section{Analytically continued gravimagnetic dipole is a wormhole}

Let us now carry out the analytical continuation $\alpha_- \to i\mu$ ($\mu>0$). The distance functions (\ref{Rr}) become
 \be\lb{Rrmu}
R_\pm = \sqrt{\rho^2+(z\pm\alpha_+)^2}, \quad r_\pm = \sqrt{\rho^2+(z\pm i\mu)^2}.
 \ee
After an overall rescaling by a factor $i$, the auxiliary functions $A, B, G$ analytically continued from (\ref{AB1}) and  (\ref{G1}) become
\ba\lb{ABG}
A_R &=& \alpha_+\mu\left[2d_+d_-(R_+R_- + r_+r_-) + (d^2+1)R_Sr_S\right] \nn\\
&& + \frac{i}2\left[d_-^2\alpha_+^2 - d_+^2\mu^2\right]R_Dr_D,\nn\\
A_I &=& 2ad\left[-\mu R_Dr_S - i\alpha_+R_Sr_D\right], \nn\\
B_R &=& 4\alpha_+\mu md\left[d_-R_S + d_+r_S\right], \nn\\
B_I &=& 4amd\left[d_-\mu R_D - id_+\alpha_+ r_D\right], \nn\\
G_R &=& 2\alpha_+\mu md\left[d_-(\alpha_+R_D - zR_S) + d_+(i\mu r_D - zr_S)\right] \nn\\
&& + m^2d\left[-d_+^2\mu R_Dr_S - id_-^2\alpha_+R_Sr_D\right], \nn\\
G_I &=& 2amd\left[\alpha_+\mu(d_-R_S + d_+r_S) \right. \nn\\
&& \left. + 2imdR_Dr_D + (-d_-\mu R_D + id_+\alpha_+r_D)z\right],
\ea
where the parameters $d$ and $a$ are now related to $m$, $\alpha_+$ and $\mu$ by
\be\lb{da}
d = \frac{\alpha_+^2 + \mu^2}{4m^2}, \quad a^2 = m^2d_+^2-\alpha_+^2 = m^2d_-^2+\mu^2,
\ee
and again $d_\pm = d\pm1$.
Taking into account the fact that $r_+$ and $r_-$ are now conjugate complex functions, and $r_D$ is now imaginary, we see that the real and imaginary parts of the analytically continued and rescaled auxiliary functions are still real. This ensures that the Ernst equations
\ba\lb{ernst}
{\rm Re}\,\E \nabla^2\E &=& (\nabla\E)^2, \nn\\
k_{,\zeta} &=& \frac12\rho f^{-2}\E_{,\zeta} \bar{\E}_{,\zeta}
\ea
(where $\zeta = \rho + iz$) which were solved by the original Ernst function are also solved by its analytical continuation, and also ensures the reality of the analytically continued metric functions. However, at first sight the metric function $e^{2k}$ given in (\ref{metfunct1}) becomes negative under the replacement of $\alpha_-^2$ by $-\mu^2$. This is easily cured by noting that the Ernst equations (\ref{ernst}) determine the metric function $k$ only up to an additive integration constant, which may be chosen equal to $i\pi/2$. This ensures that the analytically continued metric functions
\be\label{metfunct}
f = \frac{A\bar{A} - B\bar{B}}{(A+B)(\bar{A}+\bar{B})}, \quad
e^{2k} = \frac{A\bar{A} - B\bar{B}}{64d^4\alpha_+^2\mu^2R_+R_-r_+r_-}, \quad
\omega = - \frac{4{\rm Im}[G(\bar{A}+\bar{B})]}{A\bar{A} - B\bar{B}},
 \ee
yield a real Lorentzian solution of the vacuum Einstein equations.

The complex functions $r_\pm$ in (\ref{Rrmu}) are double-valued, so that the Weyl manifold $(\rho,z)$ is two-sheeted \cite{axi}.
This is made explicit by going over to oblate spheroidal coordinates $(x,\theta)$ ($0\le\theta\le\pi$),
\be\lb{oblate}
z = x\cos\theta, \quad \rho = \sqrt{x^2+\mu^2}\,\sin\theta,
\ee
leading \cite{gibbons2017} to
\be\lb{rpm}
r_\pm = x \pm i\mu\cos\theta.
\ee
The inverse transformation is double-valued, a given pair of Weyl coordinates $(\rho,z)$ corresponding either to $(x,\theta)$
or to $(-x,\pi-\theta)$. There are two Weyl sheets $(\rho,z)$, the first with $x>0$, the second with $x<0$. These two sheets are connected along the cut $x=0$. The full three-dimensional manifold is a wormhole with two locally asymptotically flat (see below) regions $x>0$ and $x<0$ connected by the two-sided disk $x=0$ ($\rho = \mu\sin\theta$, $z=0$) which is actually a topological sphere --- the wormhole throat. The successive embedded closed surfaces $x =$ constant have two poles, North ($\theta=0$), and South ($\theta=\pi$), both corresponding to $\rho=0$. Accordingly, there are two $z$-axes.

The North-South polar inversion $\theta \to \pi-\theta$ exchanges $R_+$ and $R_-$, $r_+$ and $r_-$. So the combinations $R_S$ and $r_S$ are even under this inversion, while $R_D$ and $r_D$ are odd. Inspection of (\ref{ABG}) shows that $A_R$, $B_R$ and $G_I$ are even under the inversion, while $A_I$, $B_I$ and $G_R$ are odd. It follows that the metric functions $f$, $\omega$ and $e^{2k}$ are invariant under the exchange of the two $z$-axes. On the other hand, under the exchange $x \to -x$ of the two Weyl sheets, $R_\pm$  still goes over to $R_\mp$, but now $r_\pm$ goes over to $-r_\mp$. The functions $A$, $B$ and $G$ have no simple transformation under this exchange, so that the two sides of the wormhole are not isometric.

\setcounter{equation}{0}
\section{Horizons and ergospheres}

Now we investigate the rod structure. As in the case of the gravimagnetic dipole, there are four real focal points $(\rho=0,z=\pm\alpha_+)$, two in the first sheet, and two in the second sheet. However, there are now two $z$-axes. The North $z$-axis ($\theta=0$) is divided in three rods by two focal points $x=\alpha_+$ ($z=\alpha_+$) and $x=-\alpha_+$ ($z=-\alpha_+$), the central rod $-\alpha_+<x<\alpha_+$ corresponding to a black hole horizon.  Symmetrically, the South $z$-axis ($\theta=\pi$) is divided in three rods by the two focal points $x=\alpha_+$ ($z=-\alpha_+$) and $x=-\alpha_+$ ($z=\alpha_+$), the central rod corresponding to a black hole horizon. These two black holes are not connected by a strut or Misner string, as they lie on two independent $z$-axes.

On the North horizon $\theta=0$, $-\alpha_+<x<\alpha_+$ (the South horizon will be isometric), we have
\be
R_\pm = \alpha_+ \pm x, \quad r_\pm = x \pm i\mu,
\ee
leading to $R_S=2\alpha_+$, $R_D = r_S = 2x$,  $r_D=2i\mu$, and
\ba
A_H &=& 8\mu d\left[m^2d_+(d_-\alpha_+ + d_+x) + ia(\alpha_+^2 - x^2)\right], \nn\\
B_H &=& 8\mu d\left[m\alpha_+(d_-\alpha_+ + d_+x) + iam(d_+\alpha_+ + d_-x)\right], \nn\\
G_H &=& 4\mu md\left[\alpha_+(m\alpha_+d_-^2 - \mu^2d_+) - d_+(md_+ + \alpha_+)x^2 \right.\nn\\
&& \left. + ia(d_-\alpha_+^2 - 4mdx - d_-x^2)\right].
\ea
The resulting horizon characteristics, angular velocity $\Omega_H = \omega_H^{-1}$ and surface gravity $\kappa_H = \sqrt{|e^{-2k_H}|}\,|\Omega_H|$ are:
\be\lb{omka}
\Omega_H = \frac{a}{2md_+(md_+ + \alpha_+)}, \quad \kappa_H = \frac{d\alpha_+}{2md_+(md_+ + \alpha_+)}.
\ee
The horizon area is ${\cal A}_H = 4\pi\alpha_+/\kappa_H$.

The difference
\be\lb{Delta}
\Delta \equiv A\bar{A} - B\bar{B}
\ee
being negative on the horizon rods, both $f_H$ and $e^{2k_H}$ given by (\ref{metfunct}) are negative, so that the two horizons are nested inside two symmetrical ergospheres where $f(x,\theta) < 0$ . For $x=\pm\alpha_+$, $\Delta$ vanishes for $\theta = 0$. For $x=0$ (the wormhole throat $T$), $r_\pm =  \pm i\mu\cos\theta$ and $R_\pm = R \equiv \sqrt{\alpha_+^2 + \mu^2\sin^2\theta}$, leading to
\ba\lb{throat}
A_T &=& 8\alpha_+\mu d\left[m^2d_+d_- + iaR\cos\theta\right], B_T = 8\alpha_+\mu d\left[md_-R + iamd_+\cos\theta\right], \nn\\
G_T &=& 4\alpha_+\mu md\left[(md_-^2R - d_+\mu^2)\cos\theta + iad_-R\right],
\ea
so that
\be\lb{DeltaT}
\Delta_T = (8\alpha_+\mu d)^2(a^2\sin^2\theta - \mu^2)(a^2 - \mu^2\sin^2\theta).
\ee
This is negative on the two caps $|\sin\theta| < \mu/|a|$ (recall that $\mu < |a|$ from (\ref{da})) which are the intersections of the North and South ergospheres with the wormhole throat.

\setcounter{equation}{0}
\section{Strings}

Consider now the outer rods of the North $z$-axis ($\theta=0$). For $x>\alpha_+$, we have
\be
R_\pm = x \pm \alpha_+, \quad r_\pm = x \pm i\mu,
\ee
leading to $R_S = r_S = 2x$, $R_D=2\alpha_+$, $r_D=2i\mu$, and the functions
\ba
A &=& 4\alpha_+\mu d^2\left[2x^2 - \alpha_+^2 + \mu^2 + 4m^2 \right], \nn\\
B &=& 16\alpha_+\mu md^2(x+ia), \\
G &=& - m(A+B). \nn
\ea
These result in
\be
A\bar{A} - B\bar{B} = 64d^4\alpha_+^2\mu^2(x^2-\alpha_+^2)(x^2+\mu^2).
\ee
We then find from (\ref{metfunct}) that  $\omega=0$, $e^{2k}=1$ on the two North and South outer rods $x > \alpha_+$, which are therefore regular. This regularity is inherited from that of the outer rods of the gravimagnetic dipole.

The situation is different for the ``hidden'' outer rods $\rho = 0$, $x < -\alpha_+$. On the North $z$-axis, the distance functions become
\be
R_\pm = - x \mp \alpha_+, \quad r_\pm = x \pm i\mu,
\ee
leading to $R_S = - 2x$, $r_S=2x$, $R_D=-2\alpha_+$, $r_D=2i\mu$, and
\ba
A &=& 4\alpha_+\mu\left[-2x^2 + \alpha_+^2 - \mu^2 - 4m^2d^2 \right], \nn\\
B &=& 16\alpha_+\mu md(x+ia), \\
G &=& md(A+B). \nn
\ea
From (\ref{metfunct}), we see that again
\be
\omega = 0
\ee
on the hidden outer rods, but now
\be
A\bar{A} - B\bar{B} = 64\alpha_+^2\mu^2(x^2-\alpha_+^2)(x^2+\mu^2),
\ee
so that
\be
e^{2k} = \frac1{d^4}.
\ee
These are the characteristics of semi-infinite non-rotating straight cosmic strings, with tension $(1-e^{-k_S})/4 = -d_+d_-/4$, negative for $d>1$, or positive for $d<1$.  The presence of these cosmic strings means that the metric, which is asymptotically flat for $x\to +\infty$, is only locally asymptotically flat for $x\to -\infty$.

Note that, the metric function $k$ being defined only up to an additive integration constant $k_0$, we can vary the tension of these cosmic strings, and at the same time introduce balancing cosmic strings extending to $x \to +\infty$, by varying this constant. For instance, for the choice $k_0 = 2\ln d$, the rod $x < -\alpha_+$ will now be regular, while the rod $x > \alpha_+$ will be a cosmic string with tension $(1 - d^{-2})/4 = d_+d_-/4d^2$.

\setcounter{equation}{0}
\section{Ring singularity}

Now we inquire about a possible naked ring singularity, which could show up as a zero of $A+B$, and thus a pole of the metric function $f$. The most probable location of such a singularity would be the equatorial plane $\theta = \pi/2$. Actually, in the present wormhole context, there are two equatorial planes $E_+$ ($\theta = \pi/2$, $x>0$) and $E_-$ ($\theta = \pi/2$, $x<0$). In these planes , $z=0$, $\rho = \sqrt{x^2+\mu^2}$, leading to
\be
R_\pm = R \equiv \sqrt{x^2 + 4m^2d}, \quad r_\pm = x,
\ee
and
\ba\lb{equ}
A &=& 2\alpha_+\mu(d_-R + d_+x)(d_+R + d_-x), \nn\\
B &=& 8\alpha_+\mu md(d_-R + d_+x), \\
G &=& 4ia\alpha_+\mu md(d_-R + d_+x). \nn
\ea
Accordingly,
\be
f = \frac{A-B}{A+B} = \frac{d_+R + d_-x - 4md}{d_+R + d_-x + 4md}
\ee
does not have a pole (the denominator is positive for all real $x$). However it has a double zero for $x = - md_-$, corresponding to $R = md_+$, $\rho = a$. For this value of $x$, the common factor $d_-R + d_+x$ of $A$, $B$ and $G$ also vanishes, so that $f^{-1}e^{2k}$ vanishes while $\omega$ diverges, and the metric degenerates.

Now we show that the ring $x = - md_-$, $\theta = \pi/2$ is a naked singularity. Consider the behavior of the first integrated geodesic equation
\be\lb{geo}
{\dot\rho}^2 + {\dot z}^2 + (L + E\omega)^2f^2e^{-2k}\rho^{-2} - E^2e^{-2k} = \varepsilon fe^{-2k}
 \ee
(where $\dot{}$ denotes the derivative with respect to an affine parameter, $E$ and $L$ are the energy and angular momentum of a test particle, and $\varepsilon = +1$, $0$ or $-1$ for spacelike, null or timelike geodesics)  near $x = - md_-$. Putting
$x = -m(d_- - d_+u)$, we find $R = m[d_+ - d_-u + (2d/d_+)u^2] + {\rm O}(u^3)$ for small $u$, leading to the behaviors
\be\lb{divring}
\omega = {\rm O}(u^{-2}), \quad e^{2k} = {\rm O}(u^{4}), \quad f = {\rm O}(u^{2}).
\ee
The leading terms of the effective potential in (\ref{geo}) apparently diverge as $u^{-4}$. However,
\be
\frac\rho{\omega f} = - \frac{(A+B)\rho}{4iG} = - \frac\rho{a} + {\rm O}(u^2)  = -\left(1 - \frac{m^2d_+d_-}{a^2}u\right) + {\rm O}(u^2).
\ee
So the leading divergences, of order $u^{-4}$, exactly compensate,
\be\lb{compring}
E^2\left(\frac{\omega^2 f^2}{\rho^2} -1\right)e^{-2k} \approx \frac{2E^2m^2d_+d_-}{a^2}u\,e^{-2k}
\ee
is of order $u^{-3}$, and the geodesic equation (\ref{geo}) reduces, for equatorial geodesics approaching the ring $x = - md_-$, $\theta = \pi/2$, to
\be
{\dot{u}}^2 + \frac{E^2d_+d_-}{8m^2}u^{-3} \approx 0 ,
\ee
showing that geodesics end on this ring.

\setcounter{equation}{0}
\section{Chronosphere}

It also follows from (\ref{compring}) and (\ref{divring}) that
\be
g_{\varphi\varphi} = f^{-1}\rho^2 - f\omega^2
\ee
diverges as $u^{-1}$ on the singular ring, and so must be negative on part of the equatorial plane. More precisely, for $x = -m(d_- - d_+u)$ ($|u| \ll 1$), $f \simeq u^2/4$, so that
\be\lb{signring}
g_{\varphi\varphi} \simeq -8m^2d_+d_-\,u^{-1},
\ee
is negative for $d_-u > 0$. By continuity, it must therefore also be negative on a bounded volume of the wormhole spacetime, the chronosphere, which contains closed timelike curves. Putting $f =\Delta/\Sigma$, $\omega = N/\Delta$, where $\Delta$ has been defined in (\ref{Delta}) and
\be\lb{SigmaN}
\Sigma = |A+B|^2, \quad N = 4[G_R(A_I+B_I) - G_I(A_R+B_R)],
\ee
we can write
\be
g_{\varphi\varphi} = \frac{\Sigma^2\rho^2 - N^2}{\Delta\Sigma}.
\ee
$\Sigma$ being positive, $g_{\varphi\varphi}$ is of the sign of $\Sigma\rho - |N|$ as long as $\Delta$ is positive (that is, outside the ergosphere where $\Delta < 0$, and excepting the singular ring where $\Delta = 0$).

For $\theta=\pi/2$, $x=0$ (the equator of the throat),
\be
\Sigma\rho - |N| = 64\alpha_+^2\mu^2m^4d^2d_-^2(\sqrt{d}+1)^2[(\sqrt{d}+1)^2\mu - 4\sqrt{d}|a|],
\ee
and we show in Appendix A that the last factor is negative definite. It follows that $g_{\varphi\varphi}$ is also negative on the throat equator. So we expect that for $\theta=\pi/2$,
\ba
{\rm if} \;\; d_- > 0, \;\;  & g_{\varphi\varphi} < 0 \;\; {\rm for} \;\; & -md_- < x \le 0, \nn\\
{\rm if} \;\; d_- < 0, \;\;  & g_{\varphi\varphi} < 0 \;\; {\rm for} \;\; & 0 \le x < -md_- .
\ea
By continuity, it must also be negative on the other equatorial plane in a neighborhood of $x=0$, up to a circle $x = x_0$, with sign$(x_0)$ = sign$(d_-)$. However, for ($\theta=\pi/2$, $x=+md_-$), we have again $R = md_+$, $\rho = |a|$, leading to
\be
\Sigma\rho - |N| = 64\alpha_+^2\mu^2m^4|a|d_+^4d_-^4 > 0.
\ee
The conclusion is that for $\theta=\pi/2$, if $d_- > 0$, $g_{\varphi\varphi} < 0$ for $-md_- < x < x_0 < +md_-$, and if $d_- < 0$, $g_{\varphi\varphi} < 0$ for $+md_- < x_0 < x < -md_-$.

Because $g_{\varphi\varphi}$ is negative on the equator $\theta=\pi/2$ of the throat $x=0$, it must also by continuity stay negative in a sector $\pi-\theta_0 < \theta < \theta_0$ of the throat for some $\theta_0 \in ]0,\pi/2[$. On the other hand, we show in Appendix B that $g_{\varphi\varphi} > 0$ on the throat ergocircles $\sin\theta = \mu/|a|$ where $f$ vanishes, so that $\theta_0 < \arcsin(\mu/|a|)$. We conclude that the chronosphere is a topological torus which does not intersect the North and South ergospheres.

\setcounter{equation}{0}
\section{Komar masses, angular momenta, and first law}

Let us now evaluate the horizon Komar mass and angular momenta, using the Tomimatsu approach \cite{tom83}, which we shall adapt to the present wormhole global topology. The total Komar mass and angular momentum of a stationary axisymmetric configuration
are given by the integrals over the boundary surface at spacelike infinity $\Sigma_\infty$:
 \be\lb{koMJ}
M_\infty = \frac1{4\pi}\oint_{\Sigma_\infty}D^\nu k^{\mu}d\Sigma_{\mu\nu}, \quad
J_\infty = -\frac1{8\pi}\oint_{\Sigma_\infty}D^\nu m^{\mu}d\Sigma_{\mu\nu},
 \ee
where $k^\mu = \delta^\mu_t$ and $m^\mu = \delta^\mu_\varphi$ are
the Killing vectors associated with time translations and rotations
around the $z$-axis, and $D^\nu = g^{\nu\rho}D_\rho$, with $D_\rho$ the covariant derivative. Because the integrand $D^\nu k^{\mu}$ is antisymmetric, one can
apply the Ostrogradsky theorem to transform
 \be\lb{balkom}
M_\infty = \sum_n\frac1{4\pi}\oint_{\Sigma_n} D^\nu k^\mu d\Sigma_{\mu\nu}
+ \frac1{4\pi}\int D_\nu D^\nu k^\mu dS_\mu, \lb{koM1}
 \ee
where $\Sigma_n$ are spacelike surfaces bounding the various
sources, and the second integral is over the bulk. Using again the
fact that $k$ is a Killing vector and the Einstein equations, we obtain
 \be\lb{bulk}
D_\nu D^\nu k^\mu = -[D_\nu,D^\mu]k^\nu = - {R^\mu}_\nu k^\nu =
-8\pi{T^\mu}_\nu k^\nu,
 \ee
with ${T^\mu}_\nu$ the matter energy-momentum tensor.

In the present case, $\Sigma_\infty$ has two connected components, $\Sigma_{+\infty}$ for $x\to+\infty$ and $\Sigma_{-\infty}$ for $x\to-\infty$. The bulk contribution vanishes because we are considering a vacuum solution. And the sources are the defects along the two $z$-axes --- the North and South horizons (non-rotating cosmic strings have vanishing Komar masses). So equation (\ref{balkom}) transforms into the balance equation for the Komar masses
\be\lb{balm}
M_{+\infty} + M_{-\infty} = M_{H_N} + M_{H_S}.
\ee
The same reasoning leads to the the balance equation for the Komar angular momenta
\be\lb{balj}
J_{+\infty} + J_{-\infty} = J_{H_N} + J_{H_S}.
\ee

The Komar masses and angular momenta at infinity, which coincide with the corresponding ADM observables, may be evaluated from the asymptotic behaviors of the metric functions $f$ and $\omega$. For $x\to\pm\infty$,
\be
R_\pm = |x| \pm \epsilon\alpha_+\cos\theta + {\rm O}(x^{-1}), \quad r_\pm = \epsilon|x| \pm i\mu\cos\theta,
\ee
where $\epsilon = {\rm sign}(x)$, leading to
\ba
A &=& 4\alpha_+\mu[d^2 - 1 + \epsilon(d^2 + 1)]x^2 + {\rm O}(1), \; B = 8\alpha_+\mu md(\epsilon d_+ +  d_-)|x| + {\rm O}(1), \nn\\
G &=& 4\alpha_+\mu md(d_+ + \epsilon d_-)(-|x|^2\cos^2\theta + ia\epsilon|x|\sin^2\theta) + {\rm O}(1).
\ea
From these we obtain
\ba\lb{mjinf}
M_{+\infty} = 2m,\, \quad M_{-\infty} = 2md, \nn\\
J_{+\infty} = 2ma, \quad J_{-\infty} = 2mad.
\ea
Note that, the masses at the two wormhole ends have the same sign. This contrasts with the case of other wormhole solutions, such as the Brill wormhole \cite{clement2015}, for which the masses have opposite signs at the two wormhole ends. The difference between those two masses reflects the presence of the two cosmic strings extending from the two horizons to $x\to-\infty$.

Following Tomimatsu \cite{tom83}, we evaluate the Komar horizon masses as
 \ba\lb{mh1}
M_H &=& \frac1{8\pi}\int_{H}\sqrt{|g|}g^{\rho\rho}g^{ta}\partial_\rho g_{ta}\,dzd\varphi\nn\\
&=& \frac1{8\pi}\int_{H}\left[\rho f^{-1}\partial_\rho f + \rho^{-1}f^2\omega\partial_\rho\omega\right]dzd\varphi \nn\\
&=& \frac1{8\pi}\int_{H}\left[\rho f^{-1}\partial_\rho f + \omega\partial_z\chi \right]dzd\varphi,
 \ea\
where the imaginary part of the Ernst potential (\ref{ernstpot}), $\chi = {\rm Im}\E$, is related to $\omega$ by the duality relation
 \be\lb{twist}
\partial_i\chi = -F^2\rho^{-1}\epsilon_{ij}\partial_j\omega.
\ee
The first term of (\ref{mh1}) vanishes on the horizon $\rho=0$, and $\omega$ is constant over the horizon, so we are left with
\be\lb{mh2}
M_H = \frac{\omega_H}4\left[\chi(\alpha_+) - \chi(-\alpha_+)\right].
\ee
Similar steps lead \cite{tom83,smarrnut} to the expression for the Komar horizon angular momenta
 \ba\lb{jh}
J_H &=& -\frac1{16\pi}\int_H\sqrt{|g|}g^{\rho\rho}g^{ta}\partial_\rho g_{\varphi a}\,dzd\varphi \nn\\
&=& \frac1{16\pi}\int_{H}\left[-2\omega(1-\rho f^{-1}\partial_\rho f) + \rho^{-1}(\rho^2+f^2\omega^2)\partial_\rho\omega\right]dzd\varphi \nn\\ &=& \frac{\omega_H}2\left(M_H - \alpha_+\right).
\ea

Using
\ba
A(\alpha_+) &=& 16\alpha_+\mu m^2d^2d_+, \quad B(\alpha_+) = 16\alpha_+\mu md^2(\alpha_+ + ia), \nn\\
A(-\alpha_+) &=& -16\alpha_+\mu m^2dd_+, \quad B(\alpha_+) = 16\alpha_+\mu md(-\alpha_+ + ia),
\ea
and the value of $\omega_H = \Omega_H^{-1}$ given by (\ref{omka}), we obtain the two equal Komar horizon masses and angular momenta
\be\lb{mjh}
M_{H_N} = M_{H_S} = md_+, \quad J_{H_N} = J_{H_S} = amd_+.
\ee
In agreement with (\ref{balm}) and (\ref{balj}), these add up to the sum of the Komar masses and angular momenta at $x\to\pm\infty$
given by (\ref{mjinf}). Let us note that the rotation parameter (angular momentum to mass ratio) is the same for the black holes and for the two wormhole ends,
\be
\frac{J_H}{M_H} = \frac{J_{\pm\infty}}{M_{\pm\infty}} = a.
\ee
Also, the dimensionless ratio
\be
\frac{J_H}{M_H^2} = \frac{a}{md_+} < 1,
\ee
from (\ref{da}), as in the case of the Kerr black hole.

Inverting the relation (\ref{jh}), and using (\ref{omka}), we check that the various horizon observables obey the Smarr relation
 \be\lb{smarri}
M_H = 2\Omega_H J_H + \frac{\kappa_H}{4\pi}\A_H .
 \ee
We have seen in Section 4 that the two non-interacting black holes are attached to two semi-infinite cosmic strings extending to $x\to-\infty$. So we expect that, rather than the standard first law for black hole mechanics, each  system constituted by either twin black hole and the attached cosmic string would obey the
generalized first law for a system of collinear black holes $H_n$ and cosmic strings $S_n$
 \be\lb{dyfirst}
dM = \sum_n\left(\Omega_{H_n}dJ_{H_n} + \frac{\kappa_{H_n}}{4\pi} d\A_{H_n} - \lambda_{S_n}d\mu_{S_n}\right),
 \ee
first proven for a system of two black holes in \cite{herdeiro2010}, generalized to a collinear system of many Schwarzschild black holes in \cite{gregory}, and extended to a collinear system of rotating dyonic black holes in \cite{multifirst}. In (\ref{dyfirst}),
\be\lb{strut}
\mu_{S_n} = -\frac14(1-e^{-k_{S_n}})
\ee
is the positive strut tension \cite{krtous}, opposite to the string tension, and
\be\lb{lambdaS}
\lambda_{S_n} = L_{S_n}\,e^{k_{S_n}}
\ee
(with $L_{S_n}$ the string rod length) is the string thermodynamic length \cite{appels}. In the present case, $M$ is the half total mass at infinity, equal to $M_H$, so that (\ref{dyfirst}) reads
\be\lb{first}
dM_H = \Omega_H dJ_H + \frac{\kappa_H}{4\pi} d\A_H - \lambda_S d\mu_S.
\ee

Varying independently three of the parameters, for instance $m$, $\alpha_+$ and $d$, we check that this system obeys the generalized first law (\ref{first}), with
\be\lb{mulambdas}
\mu_S = -\frac14(1-e^{-k_S}) = \frac{d^2-1}4, \quad \lambda_S = - \alpha_+ e^{k_S} = - \frac{\alpha_+}{d^2},
\ee
where the factor $-\alpha_+$ in the definition of the string thermodynamic length is the finite part of the divergent string rod length $L_S =  -\alpha_+ + \Gamma$ (with $\Gamma$ an infrared cutoff).

\setcounter{equation}{0}
\section{Limits}

\subsection{Limit $\alpha_+\to0$}

In this limit, the two horizon rods reduce to the North and South poles of the wormhole throat, corresponding to two extreme black holes. The solution is obtained from (\ref{ABG}) by expanding in powers of $\alpha_+$, with $R_{\pm} = R \pm \alpha_+ z/R + {\rm O}(\alpha_+^2)$, where $R^2 = \rho^2 + z^2$, rescaling $A = \alpha_+\hat{A}$, etc., and taking the limit $\alpha_+ \to 0$, leading to
\ba\lb{ABGextr}
\hat{A}_R &=& 2\mu\left[d_+d_-(R^2 + r_+r_-) + (d^2+1)Rr_S\right] \nn\\
&& -id_+^2\mu^2\frac{z}R r_D,\nn\\
\hat{A}_I &=& 4ad\left[-\mu\frac{z}R r_S - iRr_D\right], \nn\\
\hat{B}_R &=& 4\mu md\left[2d_-R + d_+r_S\right], \nn\\
\hat{B}_I &=& 4amd\left[2d_-\mu \frac{z}R - id_+ r_D\right], \nn\\
\hat{G}_R &=& 2\mu md\left[-2d_-zR + d_+(i\mu r_D - zr_S)\right] \nn\\
&& + 2m^2d\left[-d_+^2\mu \frac{z}R r_S - id_-^2Rr_D\right], \nn\\
\hat{G}_I &=& 2amd\left[\mu(2d_-R + d_+r_S) + 4imd\frac{z}Rr_D\right. \nn\\
&& \left.  + (-2d_-\mu \frac{z}R + id_+r_D)z\right].
\ea
This solution depends only on two parameters, for instance $m$ and $d$, the rotation parameter $a$ and the wormhole characteristic length $\mu$ being related to these by
\be
a = md_+, \quad \mu = 2m\sqrt{d}.
\ee
The general properties of this spacetime are similar to those discussed above. The two extreme black hole horizons sit on the North and South poles of the wormhole throat $x=0$. Their surface gravities vanish, while their angular velocity, area, Komar mass and angular momentum take the simple values
\be
\Omega_H = \frac1{2a}, \quad M_H = a, \quad J_H = a^2,
\ee
which are identical to those of the extreme Kerr black hole. The horizon area is $\A_H = 8\pi a^2/d$.
These horizons are surrounded by two polar caps, ergoregions where $g_{tt} > 0$. By continuity, these caps extend in the neighboring surfaces $x =$ constant to annular ergoregions.

\subsection{Limit $d \to 1$}

When $d$ goes to $1$ ($d_-\to0$), the tension $-d_+d_-/4$ of the cosmic strings along the rods $x < - \alpha_+$ goes to zero, while the area of the equatorial chronoregions where $g_{\varphi\varphi} < 0$ shrinks as $d_-^2$. Therefore we expect that the metric becomes regular in this limit.

Actually, for $d=1$, the relation (\ref{da}) between the parameters $a$ and $\mu$ reduces to $a^2 = \mu^2$, i.e. $a = \pm\mu$. Inserting this constraint together with $d=1$ in the auxiliary functions (\ref{ABG}), we find that these reduce to
\ba\lb{ABGd1}
A &=& 4\mu(\alpha_+R_S - iaR_D)\,r_\pm, \nn\\
B &=& 8\alpha_+\mu m\,r_\pm, \nn\\
G &=& - 8\mu m[\alpha_+(z-ia) + mR_D]\,r_\pm.
\ea
The complex functions $r_\pm$ , which were responsible for the two-sheeted structure, factor out completely between the numerators and denominators of the metric functions (\ref{metfunct}), which reduce \cite{manko2009} to those of the well-known Kerr black hole.

Expanding the functions (\ref{ABGd1}) in powers of $\alpha_+$, dividing by $\alpha_+$ and then taking the limit $\alpha_+ \to 0$, or equivalently taking the limit $d\to1$ in (\ref{ABGextr}), we obtain the rescaled auxiliary functions
\ba
\hat{A} &=& 8\mu\left(R - ia\frac{z}R\right)\,r_\pm, \nn\\
\hat{B} &=& 8\mu m \,r_\pm, \nn\\
\hat{G} &=& -8\mu m\left(z - ia + 2m\frac{z}R\right)\,r_\pm.
\ea
These similarly lead to the metric functions of the extreme Kerr black hole.

\section{Discussion}

We have presented and discussed a novel type of Lorentzian wormhole, a dressed wormhole in which two antipodal co-rotating black holes are embedded. The two twin black holes are hung from one of the two spacelike infinities by two semi-infinite cosmic strings with positive or negative tension. The metric also admits a naked ring singularity for almost all values of the real parameters $m$, $a$ and $d$.

This exact solution to the vacuum Einstein equations is obtained by analytical continuation of the gravimagnetic dipole vacuum solution, which features a Misner string connecting two black holes, surrounded by a chronosphere, a region containing closed timelike curves. In the case of our dressed wormhole, the two black holes, which lie on two independent North and South $z$-axes, cannot be connected by a Misner string. However there remains a toroidal region containing timelike azimuthal circles --- a Cheshire cat's grin legacy of the original Misner strings. The fact that this region is bounded does not mean that chronology violation is harmless,or confined to that region. A full investigation of the causal structure of this spacetime lies outside the limits of the present work.

We have also evaluated the Komar masses and angular momenta of the two black holes, and checked that these obey the usual Smarr relation, as well as the generalized first law for a system of collinear black holes and cosmic strings.

A drastic simplification occurs when the dimensionless parameter $d$ takes the special value $d=1$. In this case the cosmic strings disappear (their tension vanishes), and geodesics no longer end on the equatorial ring $\rho = a$. Actually, the metric reduces in this case to that of the well-known Kerr black hole \cite{manko2009}.

When the real parameter $\alpha_+$ goes to zero, the two black holes become extreme, their horizons sitting on the North and South poles of the wormhole throat. One may speculate what would happen if $\alpha_+$ were then continued to imaginary values $i\nu$. Both distance functions $R_\pm$ and $r_\pm$ would then be complex, leading to a four-sheeted Weyl manifold. The resulting three-dimensional geometry would presumably correspond to a wormhole with multiple throats \cite{makita,herr}.

The solution discussed in this paper is given in Weyl coordinates, which are such that black hole horizons are mapped onto segments of the $z$-axes, and so cover only the stationary domain. The analytical extension of this solution beyond the black hole horizons deserves further investigation.

There is no reason why this new dressed wormhole should be unique. We suggest that other dressed wormhole spacetimes could be generated by a similar analytic continuation technique from known multi-black hole spacetimes, for instance as solutions to the coupled Einstein-Maxwell equations.

\section*{Acknowledgment}

I thank D. Gal'tsov for enlightening comments.

\renewcommand{\theequation}{A.\arabic{equation}}
\setcounter{equation}{0}
\section*{Appendix A}

Let us show that the quantity
\be
(\sqrt{d}+1)^2\mu - 4\sqrt{d}|a|
\ee
is negative definite. This is of the same sign as
\ba
(\sqrt{d}+1)^4\mu^2 - 16d\,a^2 &=& [(\sqrt{d}+1)^4 - 16d]\mu^2 - 16dd_-^2m^2 \nn\\
&=& (\sqrt{d}-1)^2(d+6\sqrt{d}+1)\mu^2 - 16d(\sqrt{d}+1)^2m^2 \nn\\
&<& - (\sqrt{d}-1)^2(3d + 2\sqrt{d} + 3)\mu^2 < 0.
\ea

\renewcommand{\theequation}{B.\arabic{equation}}
\setcounter{equation}{0}
\section*{Appendix B}

Let us show that $g_{\varphi\varphi}$ is positive on the throat ergocircles $x=0$, $\sin\theta = \mu/|a|$ where $g_{tt}$ vanishes. On the throat, from (\ref{throat}), $N$ and $\Sigma$ defined by (\ref{SigmaN}) take the values
\ba
N &=& -2(8\alpha_+\mu d)^2(md_++R)(md_-^2R\sin^2\theta + \mu^2d_+\cos^2\theta), \nn\\
\Sigma &=& (8\alpha_+\mu d)^2(md_++R)^2(m^2d_-^2 + a^2\cos^2\theta).
\ea
So
\ba
\Sigma\rho - |N| &=& (8\alpha_+\mu d)^2(md_++R)[\mu(md_++R)((m^2d_-^2 + a^2\cos^2\theta)\sin\theta \nn\\
&& - 2|a|m(md_-^2R\sin^2\theta + \mu^2d_+\cos^2\theta)].
\ea
This vanishes on the ergocircle $\sin\theta = \mu/|a|$. Near the ergocircle, for $a\sin\theta = \mu + au$ ($u \ll 1)$, we find
\be
\Sigma\rho - |N| \approx (8\alpha_+\mu d)^2(m^2d_+^2 - R^2)\mu a^2\,u,
\ee
while, from (\ref{DeltaT}),
\be
\Delta \approx (8\alpha_+\mu d)^2(m^2d_+^2 - R^2)2\mu |a|\,u.
\ee
Hence,
\be
g_{\varphi\varphi} = \frac{\Sigma^2\rho^2 - N^2}{\Delta\Sigma} = \frac{2\rho(\Sigma\rho - |N|)}{\Delta} = \mu^2.
\ee

\end{document}